\documentstyle[12pt]{article}

\begin{document}

\title{{\bf Taub's plane-symmetric vacuum spacetime revisited}}
\author{
M. L. Bedran\thanks{Universidade Federal do Rio de Janeiro, Instituto
de F\'{\i}sica, Caixa Postal 68528, CEP 21945-970, Rio de Janeiro~--~RJ
Brazil}\ \thanks{\sc internet: bedran@if.ufrj.br}\ ,
\ \ 
M. O. Calv\~ao$^\ast$\thanks{\sc internet: orca@if.ufrj.br}\ ,
\ \
I. Dami\~ao Soares\thanks{Centro Brasileiro de Pesquisas F\'\i sicas, R. Dr.
Xavier Sigaud 150, 22290-180 Rio de Janeiro~--~RJ, Brazil, {\sc internet:
ivanods@lafex.cbpf.br}}
\\ \ \ and\ \
F. M. Paiva\thanks{Departamento de Astrof\'{\i}sica, Observat\'orio
Nacional~--~CNPq, Rua General Jos\'e Cristino 77, 20921-400 Rio de
Janeiro~--~RJ, Brazil, {\sc internet:  fmpaiva@on.br}}
}

\date{\today}

\maketitle

\begin{abstract}
\noindent The gravitational properties of the {\em only\/}
static plane-symmetric vacuum solution of Einstein's field
equations without cosmological term (Taub's solution, for
brevity) are presented: some  already known properties
(geodesics, weak field limit and pertainment to the
Schwarzschild family of spacetimes) are reviewed in a physically
much more transparent way, as well as new results about its
asymptotic structure, possible matchings and nature of the
source are furnished.  The main results point to the fact
that the solution must be interpreted as representing the
exterior gravitational field due to a {\em negative\/} mass
distribution, confirming previous statements to that effect in the
literature.  Some analogies to Kasner's spatially homogeneous
cosmological model are also referred to.

\end{abstract}
{\sc pacs} numbers: 04.20.Jb, 04.20.Cv

\section{Introduction}  \setcounter{equation}{0}

The main aim of this paper is to provide a detailed interpretative
investigation of the vacuum plane-symmetric spacetimes of general
relativity. Some generic motivating reasons may be adduced for that.

Firstly, the presence of so-called topological defects
(monopoles, cosmic strings, domain walls and textures), arising
from  cosmological phase transitions (symmetry breaking), have
become quite fashionable in the study of the very early
universe, particularly as concerns their deep consequences for
large-scale structure formation and cosmic background radiation
anisotropies\cite{Peebles1993}-\cite{Coulsonetal1996}.
All these objects have peculiar gravitational properties;
generic walls or shells (and also textures, to that effect),
however, seem to have a slightly more acceptable (mathematical)
behavior in the sense that, contrarily to one- or
two-dimensional trapped regions, their spacetimes can be
described by curvature tensors well defined as mathematical
distributions\cite{Gerochetal1987}.

Secondly, the Casimir effect associated with the vacuum bounded by two
infinite parallel plates, when treated in a fully consistent way,
should take into account the proper gravitational field of the
plates \cite{Birrelletal1982}. In this setting, it turns out to be an
issue for quantum field theory in curved spacetime.

Thirdly, when trying to formalize, in a more precise way, the
equivalence between inertial and gravitational effects, the
problem of what the general relativistic
model of a spatially homogeneous gravitational field is has to
be faced. Its resolution, from a naive Newtonian point of view, suggests
again the paying attention to the vacuum plane-symmetric
solutions of Einstein's field equations. Of course, the
determination of the Newtonian gravitational field outside a
static, infinite, uniformly dense planar slab is a trivial
exercise; however, as we shall see, the general relativistic
situation is not so plain.

The paper is organized as follows. In Sec.~\ref{sec:psv}, we expose the
properties of Taub's  plane-symmetric vacuum model:  isometries,
singularity, kinematics of the observers adapted to the
symmetries, timelike and null geodesics (including geodesic
deviation for timelike ones). These aspects have already been
dealt with in the
literature\cite{Taub1951}-\cite{Bonnor1990};
however, not only for the sake of completeness and fixing
conventions, but also for further clarification and
generalization do we delve again into them.  In
Sec.~\ref{sec:boundaries}, we study the asymptotic structure as
well as the Newtonian limit. In Sec.~\ref{sec:junction}, we
prove that Taub's global solution is the limit of the
mirror-symmetric matching of two Taub domains, to the ``left''
and ``right'' of a negative mass planar shell.  In
Sec.~\ref{sec:discussion}, we discuss the main implications of
our results and present a conclusion. We also provide an
Appendix where the plane-symmetric vacuum solutions are shown,
via Cartan's invariant technique, to be parametric limits of a
generalized Schwarzschild family of spacetimes.

The signature of the metric, except for Sec.~\ref{sec:junction}, is
$-2$; also $c=8\pi G=1$.

\section{Plane-symmetric vac\-u\-um models}  \setcounter{equation}{0}
\label{sec:psv}

\subsection{Metrics}
\label{sec:metrics}

Einstein's {\em vacuum} field equations, without cosmological
term, for a plane-symmetric geometry, will be satisfied by only
three distinct {\em single\/} (without any essential constants
of integration) solutions\cite{Krameretal1980}: Taub's static
metric, Kasner's spatially homogeneous, and Minkowski metric.
\begin{itemize}
\item {\sc taub's geometry (t):}
\begin{eqnarray}
     ds^2_{(T)} & = & \frac{1}{z^{2/3}}dt^2-z^{4/3}\left(dx^2+
                              dy^2\right) -dz^2,    \label{Taub} \\
                & = & \frac{1}{r}dt^2-r^2\left(dx^2+dy^2\right)-rdr^2,
                                                      \label{Taub-r}
\end{eqnarray}
\noindent with coordinate ranges
\begin{equation}
      -\infty<t,x,y<+\infty,\ 0<r<+\infty\ \ \ \mbox{and}
               \left\{  \begin{array}{l}
                       0<z< +\infty,\ \mbox{or} \\
                       -\infty<z<0
               \end{array}
               \right.                         ;
                                                 \label{Taubcoord}
\end{equation}

\item {\sc kasner's geometry (k):}
\begin{eqnarray}
     ds^2_{(K)} & = & dt^2-t^{4/3}\left(dx^2+dy^2\right) - 
                           \frac{1}{t^{2/3}}dz^2,  \label{Kasner} \\
                & = & tdt^2-t^2\left(dx^2+dy^2\right)-\frac{1}{t}dr^2,
                                                      \label{Kasner-r}
\end{eqnarray}
\noindent with coordinate ranges
\begin{equation}
      -\infty<x,y,z,r<+\infty\ \ \ \mbox{and}\
                                \left\{ \begin{array}{l}
                             0<t<+\infty,\ \mbox{or} \\
                            -\infty<t<0
                         \end{array}
                            \right.        ; 
                                                \label{Kasnercoord}
\end{equation}

\end{itemize}

The similarity of Taub's and Kasner's geometries to Rindler's
and Milne's universes, respectively, is to be noticed.

\subsection{Singularities and isometries}
\label{sec:singularities}

The coordinate ranges for Taub's and Kasner's models given
above are those where the solutions are regular.  As a matter of
fact, the only nonvanishing Carminati-McLenaghan algebraic
invariants\cite{Carminatietal1991} for Taub's and Kasner's
solutions are:
\begin{equation}
       I_{1(T)}  = \frac{64}{27z^4},\quad
       I_{2(T)} = -\frac{256}{243z^6},      \label{invTaub}
\end{equation}
\noindent and
\begin{equation}
       I_{1(K)} = \frac{64}{27t^4},\quad
       I_{2(K)} = \frac{256}{243t^6},      \label{invKasner}
\end{equation} 
\ \\
\noindent where $I_1:=R^{\alpha\beta\gamma\delta}
R_{\alpha\beta\gamma\delta}$ 
and $I_2:= {R_{\alpha\beta}}^{\mu\nu}{R_{\mu\nu}}^{\lambda\rho}
{R_{\lambda\rho}}^{\alpha\beta}$. 
     
As concerns Taub's solution, we call attention to the following
properties: (i)~it is {\em curved\/} ; (ii)~it is not only
plane-symmetric, but also {\em static\/}; (iii)~it has a
timelike singularity at $z=0$ ($r=0$); (iv)~its algebraic invariants
vanish when $z,r\rightarrow +\infty$ (suggesting it is
asymptotically flat at spatial infinity in the $z$ direction;
cf. Sec.~\ref{sec:boundaries}). Kasner's solution presents the
``dual'' properties: (i)~it is also {\em curved\/} ; (ii)~it is
not only plane-symmetric, but also {\em spatially homogeneous\/}
(beware: it is {\em not\/} stationary!); (iii)~it has a
spacelike singularity at $t=0$; (iv)~its algebraic invariants
vanish when $t\rightarrow +\infty$.

\subsection{Kinematics of adapted observers}
\label{sec:kinematics}

The motion of the observers adapted to those coordinate systems
$\left(u^\alpha=\right.$ $\left. =\delta^\alpha_0/\sqrt{g_{00}}\right)$ is
co\-vari\-ant\-ly char\-ac\-ter\-ized by the ki\-ne\-mat\-ic
quantities\cite{Ellis1971}: acceleration $a^\alpha$, expansion
$\Theta$, shear $\sigma^{\alpha\beta}$ and vorticity
$\omega^{\alpha\beta}$ .  For the models under consideration, we
have only the following nonvanishing quantities
\begin{equation}
      a^\alpha_{(T)}=-\frac{1}{3z}\delta^\alpha_3;
                                             \label{kpTaub}
\end{equation}

\begin{equation}
      \Theta_{(K)}=\frac{1}{t},\quad
           \sigma_{\alpha\beta\,(K)} = \mbox{diag}\left(0,
           -{\scriptstyle\frac{1}{3}}t^{1/3},
           -{\scriptstyle\frac{1}{3}}t^{1/3},
           {\scriptstyle\frac{2}{3}}t^{-5/3}\right);  \label{kpKasner}
\end{equation}

\noindent We see that the adapted observers of Taub's,
Eq.~(\ref{Taub}), solutions
constitute a {\em rigid nonrotating accelerated frame\/}, whereas
those of Kasner's, Eq.~(\ref{Kasner}),
solutions constitute a {\em deforming nonrotating geodesic frame\/}.
The results for Taub's observers are obvious since they
are manifestly static (i.e., their four-velocities are proportional
to irrotational timelike Killing vector fields of the respective
geometries), so that they remain at rest with respect to one another
and with respect to the singularity too. This is a consequence of the
physically more intuitive fact that the proper time, as measured by a
static observer, for a photon to travel to and back from a second
static observer is independent of the emission event. Half this
proper time defines the so-called  {\em radar distance\/}
between the two observers, which is a coordinate-independent concept.
Specifically, for Taub's spacetime, this radar distance between two
static observers ${\cal{O}}_1$ and ${\cal{O}}_2$ with respective
spatial coordinates $(x,y,r_1)$ and $(x,y,r_2)$, as measured by
${\cal{O}}_1$, is given by
\begin{eqnarray}
   \Delta s({\cal{O}}_1,{\cal{O}}_2) & = & \sqrt{g_{00}(r_1)}
   \left|\int_{r_1}^{r_2} \sqrt{-{g_{33}(r)}/{g_{00}(r)}}
   \,dr\right|                     \nonumber \\ & = &
   \frac{1}{2\sqrt{r_1}}|r_2^2 - r_1^2|\neq \Delta s({\cal{O}}_2,{\cal{O}}_1)\,.
\end{eqnarray}
\noindent As explicitly displayed, this radar distance is {\em not\/}
symmetric, a feature which already happens to Rindler's
observers in Minkowski spacetime and is due to the relativity of
the simultaneity for the different instantaneous inertial frames
attached to each observer. We also notice that, as the radial
coordinate distance $\Delta r({\cal{O}}_1,{\cal{O}}_2):=
|r_2-r_1|$ increases, so does the radar distance. This result
will be used in Sec.~{\ref{sec:geodesics}} to show that free
particles are ``repelled'' from the singularity.

\subsection{Geodesics} 
\label{sec:geodesics}

The Lagrangian for the geodesics of metric (\ref{Taub-r}) is

\begin{equation}
      {\cal{L}}=\frac{1}{2}\left[\frac{1}{r}{\dot t}^2-r^2\left({\dot
x}^2 + {\dot y}^2\right) - r{\dot r}^2\right],    \label{Lagrangian}
\end{equation}

\noindent where a dot denotes a derivative with respect to  an affine
parameter $\tau$ (the proper time, for timelike geodesics). We have
exactly three linearly independent Killing vector fields, so that the
corresponding canonical momenta are conserved:

\begin{equation}
             \frac{1}{r}{\dot t}=: E = const>0,     \label{energy}
\end{equation}
\begin{equation}
            r^2\dot x =: p_x = const,    \label{xmom}
\end{equation}
\begin{equation}
             r^2\dot y =: p_y = const.     \label{ymom}
\end{equation}

\noindent Furthermore, the invariance of the character of the
geodesic ($2{\cal{L}}=\epsilon:=0\mbox{\ or\ }+1$, for null or
timelike geodesics, respectively) implies now

\begin{equation}
             \dot r^2 = E^2 - V(r),     \label{charinv}
\end{equation}

\noindent with

\begin{equation}
           V(r):=\frac{1}{r^3}\left(p_{\mid\mid}^2 +\epsilon
                                 r^2\right)\geq 0    \label{effpot}
\end{equation}

\noindent and

\begin{equation} 
            p_{\mid\mid}:=\sqrt{p_x^2 + p_y^2}.       \label{parmom}
\end{equation}

The motion in the coordinate $r$ is thus reduced to a usual one-dimensional
problem with the {\em effective potential\/} (\ref{effpot}) and {\em
r coordinate\/} acceleration
\begin{equation}
      \ddot r = \frac{1}{2r^4}\left(3p_{\mid\mid}^2 + \epsilon
                           r^2\right) \geq 0.        \label{r-accel}
\end{equation}
From either (\ref{charinv})--(\ref{parmom}) or
(\ref{parmom}),(\ref{r-accel}), we see that we have two
qualitatively distinct cases, according to the mass
and$\backslash$or the initial conditions of the test particle:
(a)~massless particles in purely ``perpendicular'' motion 
($\epsilon=p_{\mid\mid}=0$); (b)~massive particles ($\epsilon=1$) or massless
particles in ``transverse'' motion ($\epsilon=0\not=p_{\mid\mid}$). Case
(a) implies $V(r)\equiv 0,\ \ddot r\equiv 0$. These null
geodesics {\em do\/} attain the singularity. Case (b) implies
$V(r)\rightarrow +\infty,\
\ddot r\rightarrow +\infty$ when $r \rightarrow 0_+$ and
$V(r)\rightarrow 0_+,\ \ddot r\rightarrow 0_+$ when $r\rightarrow
+\infty$. These geodesics can {\em never attain\/} the singularity.

In fact, {\em all} massive geodesic particles
or even generic ($p_{\mid\mid}\neq0$) massless geodesic particles are
{\em repelled} from the singularity as described by the static
observers $r=const$. This is clear from the following. As shown at
the end of Sec.~\ref{sec:metrics}, the static observers are at rest
with respect to each other and with respect to the singularity. Since
the $r$ coordinate for a geodesic particle eventually increases
without bound, it will cross the static observers receding further
and further away from the singularity. Furthermore, a geodesic
particle initially at rest will be displaced to increasing values of
$r$, as can be derived from (\ref{r-accel}). This implies that,
relative to the static observers, a geodesic particle is accelerated
away from the singularity. This description is consistent with the
negative mass interpretation for the sources of Taub's spacetime, as
discussed at the end of this subsection, in the Appendix and in
Sec.~\ref{sec:junction}.

Case (a) above can be further examined by the trivial integration of
(\ref{charinv}), taking into account that $V(r)\equiv 0$:
\begin{equation}
    r(\tau) = \left\{ \begin{array}{lcr}
                       r_{(0)}+E\tau, & \mbox{for $-r_{(0)}/E\leq
                                  \tau<+\infty;$} & \mbox{({\em
                                              outgoing})} \\
                       r_{(0)}-E\tau, & \mbox{for $-\infty<\tau\leq
                                  +r_{(0)}/E;$} & \mbox{({\em incoming}).}
                      \end{array}
              \right.
                                          \label{purperp-r}
\end{equation}        
\noindent Here we have chosen the initial condition $r_{(0)}:=r(\tau\!=\!0)$.
Replacing (\ref{purperp-r}) into (\ref{energy}), (\ref{xmom}) and
(\ref{ymom})  allows us to
determine the remaining parametric equations of these geodesics:
\begin{equation}
     t(\tau) = \left\{\begin{array}{lcr}
                   r_{(0)}E\tau+{\scriptstyle\frac{1}{2}}E^2\tau^2, &
                             \mbox{for $-r_{(0)}/E\leq \tau<+\infty;$}
                                          & \mbox{({\em outgoing})} \\
                   r_{(0)}E\tau-{\scriptstyle\frac{1}{2}}E^2\tau^2, &
                             \mbox{for $-\infty<\tau\leq
                                  +r_{(0)}/E;$} & \mbox{({\em incoming});}
                       \end{array}
               \right.
                                            \label{purperp-t}\\
\end{equation}
\begin{equation}    
     x(\tau)  =  x_{(0)};  \label{purperp-x}\\
\end{equation}
\begin{equation}    
 y(\tau)  =  y_{(0)},  \label{purperp-y}
\end{equation}
\noindent with  the choice $t(\tau\!=\!0)=0$. These null
geodesics may reach the boundaries $r=+\infty$ (to be defined in
Sec.~\ref{sec:boundaries}), but, as we see from
(\ref{purperp-r}), they take an infinite interval of affine
parameter to reach them.

 The case of massive particles
($\epsilon=1$) in purely ``perpendicular'' motion ($p_{\mid\mid}
= 0$)
is also easily dealt with as regards the (inverse) parametric equation of
motion for the coordinate $r$, leading to
\begin{eqnarray}
 \pm E^3\tau(r) & = & E\sqrt{\left(E^2r - 1 \right)r} +
\mbox{Arc\,cosh}\,\left(\sqrt{E}r \right) \nonumber \\
   &  & - E\sqrt{\left(E^2r_0 - 1 \right)r_0} -
\mbox{Arc\,cosh}\,\left(\sqrt{E}r_0 \right), \label{massiveperp}
\end{eqnarray}
where, again, $r_0 := r(\tau\!=\!0)$ and the $+$ sign corresponds to the
outgoing motion and the $-$ sign to the incoming one. These
results will be useful presently (cf. Secs.~\ref{sec:boundaries}
and \ref{sec:junction}).

To further interpret the model, let us now calculate, from the
geodesic deviation equation, the 
relative acceleration between two geodesics at instantaneous rest
with respect to the singularity ($\dot x\!\mid_0\,=\dot y\!\mid_0\,=\dot
r\!\mid_0\,=0$). We find
\[
\left.\frac{D^2\eta^0}{D\tau^2}\right|_0  =  0,
\]
\begin{equation}
\left.\frac{D^2\eta^A}{D\tau^2}\right|_0 = +\frac{1}{2r^3}\eta^A,\ \ (A=1,2),
                                      \label{geoddev}
\end{equation}
\[
\left.\frac{D^2\eta^3}{D\tau^2}\right|_0 = -\frac{1}{r^3}\eta^3.
\]
Note that the signs in the above equations are the opposite of the
corresponding ones in the Schwarzschild spacetime.
Thus, a set of two freely falling particles released from rest (as just
described above) may behave in two characteristic ways:
(a)~if their initial spatial positions have equal $x$ and $y$ but
different $r$, their relative distance will instantaneously decrease,
while being repelled from the singularity; (b)~if their initial
spatial positions have equal $r$ but different $x$ and/or $y$,
their relative distance will instantaneously increase, while
being repelled from the singularity again. Contrary to the case
of the curvature field being spatially homogeneous (independent
of $r$), as occurs in the Newtonian theory for the gravitational
field {\bf g}, the plane  symmetry of the metric field does not imply
homogeneity (neither in the metric nor the curvature fields):
analogous space-time measurements (such as those for the
geodesic deviation) at events with  different $r$
coordinates furnish distinct results.

\subsection{Parametric limits}
\label{sec:limits}

Some intuition on the nature of the source (singularity) of
Taub's (global) spacetime might also be expected to arise from a
study of the families of metrics to which it belongs. The
problem with this argument is twofold: (i)~a given metric may be
a parametric limit of several disjoint families of metrics,
(ii)~a given family of metrics may have different limits, for
the same limiting values of its parameters, when the limiting
process is carried out in different coordinate systems. This has
been first pointed out by Geroch\cite{Geroch1969}, who
presented Schwarzschild's metric in three different coordinate
systems, such that, as the mass tends to $+\infty$, the limiting
result is either singular, Minkowski's metric or Kasner's
(spatially homogeneous) metric. Due to the resemblance between
Kasner's and Taub's metrics, one naturally wonders whether the
latter is also a limit of the Schwarzschild family of metrics as
the mass tends to $+\infty$; the answer, however, is negative,
since the only Petrov type D such limits are Minkowski's and
Kasner's (spatially homogeneous) spacetimes\cite{Paivaetal1993}.
Still,  coordinate systems exist, for this Schwarzschild family
of metrics, such that, now in the limit $m\rightarrow -\infty$,
one ends up with Taub's metric\cite{Horskyetal1982,Soares1976}.
In the Appendix, we analyze a generalized Schwarzschild family
of metrics, which family is closed in the sense of Geroch
\cite{Geroch1969}. This analysis is carried out by means of Cartan
scalars, which characterize a given geometry independently of
coordinate systems, and the above results are explicitly derived in
an invariat way and also cast in a wider context.

It is expedient to mention here some other families to which Taub's metric
belongs. First, we have two subfamilies of Weyl's static vacuum
metrics\cite{Bonnor1990}: the Levi-Civita metrics and the
Parnovski-Papadopoulos metrics. Then we have the generalized
Kasner metrics\cite{Harvey1990}. Finally, also the
Robinson-Trautman vacuum solutions\cite{Krameretal1980} and the
Kerr-Schild solutions\cite{Krameretal1980}.

\section{Asymptotic structure and the weak field limit} \label{sec:boundaries} 
\setcounter{equation}{0}

In this section, we introduce coordinate systems which render
explicit the null boundary structure of the spacetime and also
characterize a weak field limit to Taub's geometry.

Let us consider the static plane-symmetric Taub metric in the form
\begin{equation}
\label{Tds2}
ds^2_T = \frac{1}{r}dt^2 -rdr^2-r^2(dx^2+dy^2)
\end{equation}
($0<r<\infty$) and introduce in the manifold of (\ref{Tds2}) the Kruskal-type
coordinate system ($u, v, x, y$) defined in Table 1.

In this coordinate system, the metric (\ref{Tds2}) assumes the form
\begin{equation} \label{ds2t}
ds^2_T= \frac{2\sqrt{2}(du^2-dv^2)}{(v^2-u^2) \sqrt{-\ln(v^2-u^2)}} +
2\ln(v^2-u^2)(dx^2+dy^2)
\end{equation}
We note that the curvature tensor of (\ref{ds2t}) tends continuously
to zero as
$v^2-u^2\rightarrow 0$. Asymptotic coordinates may be introduced such that
 (\ref{ds2t}) becomes regular on $v^2-u^2=0$ (cf.~Table 1). The boundaries
 $v^2-u^2=0$ are in fact two flat null surfaces at infinity ($r=+\infty$),
 which 
we shall denote by $J^+$ and $J^-$ (cf. Fig.~\ref{I1}), and correspond to
 the asymptotically flat regions of the spacetime.

If, in (\ref{Tds2}), we extend the domain of the coordinate $r$ to
$-\infty<r<\infty$, the resulting manifold of the geometry (\ref{Tds2})
is the union of a Kasner (K) spacetime to a Taub (T) spacetime, with
the singular locus $r=0$ as a common boundary. The Kasner spacetime
corresponds to the domain $-\infty<r<0$, its singularity $r=0$ having
now a spacelike character. A typical plane $(u,v)$ of Kasner's spacetime
is represented in Fig.~\ref{I2}. The metric in K is given by
\begin{equation} \label{ds2k}
ds^2_K= \frac{2\sqrt{2}(du^2-dv^2)}{(u^2-v^2) \sqrt{-\ln(u^2-v^2)}} +
2\ln(u^2-v^2)(dx^2+dy^2)
\end{equation}
Expression (\ref{ds2k}) is obtained by interchanging $u$ and $v$ in the
metric components of (\ref{ds2t}). Note that in the K region $0\leq
u^2-v^2\leq 1$, and $(u,v)$ are still timelike and spacelike
coordinates, respectively.

A further coordinate system is now introduced (cf. Table 1),
 which will be useful in
examining the asymptotic form  of Taub's metric. 
In the coordinate system $(\mu, \nu, x, y)$, Taub's geometry (\ref{ds2t}) 
assumes the form
\begin{equation} \label{ds2tg}
ds^2_T= -\frac{2\sqrt{2}d\mu d\nu}{\sqrt{\mu+\nu}} -
2(\mu+\nu)(dx^2+dy^2)
\end{equation}

and Kasner's geometry (\ref{ds2k}) becomes, 
\begin{equation} \label{ds2kg}
ds^2_K=  \frac{2\sqrt{2}d\mu d\nu}{\sqrt{\mu+\nu}} -
2(\mu+\nu)(dx^2+dy^2)
\end{equation}
The null infinities of Taub's geometry are now expressed by
\begin{equation}
\begin{array}{lcl}
 J^+ & : & \mu\rightarrow \infty \\
 J^- & : & \nu\rightarrow \infty 
\end{array}
\end{equation}
and analogously for Kasner's geometry. The diagrams of Fig.~\ref{I4}
make explicit the relation between the $(u,v)$ and $(\mu, \nu)$
coordinates.

Although we have introduced coordinate systems (cf.
Table~\ref{coordinates}) where the asymptotic boundaries of Taub's
geometry are characterized and explicitly exhibited (the two flat null
infinities ${J^+}$ and ${J^-}$, and the spatial infinity $i^0$),
the metric in these coordinate systems is not well defined in the
boundaries. In order to give a regular expression for  Taub's metric
in a neighborhood of the boundaries ${J^+}$ and ${J^-}$, we
therefore introduce two new sets of asymptotic coordinates (cf. Table
1).

We rewrite (\ref{ds2tg}) as
\begin{equation} \label{ds2tgl}
ds^2_T= - \frac{2\sqrt{2}d\xi d\nu}{\sqrt{1+\frac{4\nu}{\xi^2}}} -
\frac{1}{2}\xi^2(1+4\frac{\nu}{\xi^2}) (dx^2+dy^2)
\end{equation}
Near ${J^+}$ we have $4\nu/\xi^2\rightarrow 0$, and we may expand
the above line element as
\begin{equation}
ds^2_T\approx ds^2_M + h
\end{equation}
where
\begin{equation}
ds^2_M= - 2\sqrt{2} d\xi d\nu 
- \frac{1}{2}\xi^2 (dx^2+dy^2)
\end{equation}
and 
\begin{equation}
h= \frac{4\sqrt{2}\nu}{\xi^2} d\xi d\nu -
2\nu (dx^2+dy^2)
\end{equation}
We note that the line element $ds^2_M$ is the Minkowski one in
coordinates $x^\alpha = (\xi, \nu, x, y)$. Indeed, let us
realize a further coordinate transformation (cf. Table 1),
which casts $ds^2_M$ into the form
\begin{equation} \label{gmm}
ds^2_M = - d\xi dN - (dX^2+dY^2)
\end{equation}
and all components of $h$ in this new coordinate system are at least
$O({1}/{\xi^2})$, as well as its first and second derivatives.
Therefore, in the coordinate system $(\xi, N, X, Y)$ the metric near
${J^+}$ is regular, as well as its first and second order derivatives,
and differ from the Minkowski metric by a small perturbation $h$ which
tends to zero as $\xi\rightarrow\infty$.

An analogous expansion may be realized in a neighborhood of ${J^-}$ :
$\nu\rightarrow\infty$. This may be obtained from the expressions of
the above paragraph by the obvious substitution
$\nu\rightleftharpoons\mu$.

We can now discuss the weak field limit for Taub's  geometry.
Equation (\ref{gmm}) suggests the introduction of a Cartesian coordinate
system at infinity, namely, $(T, Z, X, Y)$, defined in Table 1.
Thus, when $z\rightarrow\infty$ for finite $T$, we have
$\xi\rightarrow\infty$ and $N\rightarrow\infty$, with 
$N/\xi^2\rightarrow 0$. Therefore, the region described by
$Z\rightarrow\infty$ for finite $T$ is a neighborhood of the spatial
infinity $i^0$ of Fig.~\ref{I1}. The asymptotic expression of Taub's
metric (for $X$ and $Y$ finite) is then
\begin{equation} \label{gT}
ds^2_T = dT^2 - dX^2 - dY^2 -dZ^2 + \frac{4N}{\xi^2}
\left(\frac{1}{2}(dZ^2-dT^2) - dX^2-dY^2\right),
\end{equation}
and, in a neighborhood of the spatial infinity $i^o$, (\ref{gT})
reduces to
\begin{equation} \label{gT2}
ds^2_T = (1-2/Z)(dT^2 - dZ^2) - (1+4/Z)(dX^2+dY^2)
\end{equation}
A Newtonian-like potential at $i^0$ is then given by
\begin{equation}
\Phi_N = -1/Z.   \label{Phi}
\end{equation}
If we choose a plane $Z=Z_0$ in a neighborhhod of $i^0$, we can expand
(\ref{Phi}) about $Z_0$ as
\begin{equation}
\Phi_N = -1/(Z_0+\ell) \approx -(1/Z_0)(1-\ell/Z_0)
\end{equation}
Then, near the plane $Z=Z_0$, we have a homogeneous field orthogonal to
$Z_0$ with strength $1/Z_0^2$, for all ($X, Y$) finite.

This configuration of the gravitational field (in a neighborhood of
$i^0$) is the nearest to a homogeneous gravitational field we may
achieve in the static plane-symmetric Taub geometry. In the past
literature, this problem of the nonrelativistic limit 
to a static plane-symmetric metric was considered in \cite[Sec. 
IV]{Aichelburg1970}.  The treatment there is, however, incomplete: a
coordinate system is introduced where Taub's metric is put in the form
$ds^2_T = ds^2_M + h$, where $h$ is a small (in a specific region) deviation of
the Minkowski metric, $ds^2_M$. Nevertheless, the connection and
curvature tensor components are not small (of the order of $h$)
in that region, unless he specifies that $g\rightarrow 0$, thus
fixing the asymptotic region ($z\rightarrow +\infty$) as the
region of true nonrelativistic limit; in this case, the above
coordinate system is well defined only in the asymptotic region
(cf.~\cite[Eq.~(12)]{Aichelburg1970}). Furthermore, since Taub's
metric has no essential free parameter in it, there simply does
not exist a quantity which would play the role of the mass
density of the source.

\section{Matching and shell} 
\setcounter{equation}{0}
\label{sec:junction}

We now implement a {\em mirror-symmetric\/} junction of two Taub
domains by means of a {\em negative\/} mass shell, in order to
show that Taub's original spacetime (\ref{Taub}) may be naturally viewed as
the limit of this matched one as the surface energy density of
the shell approaches $-\infty$. To this end, we shall take
advantage of Israel's shell formalism \cite{Israel1966},
following all his conventions inclusively; in particular, the
signature of the metric, in this section, is +2.

The four-dimensional manifolds $V^+$ and $V^-$ will be taken as
Taub domains; specifically, $V^+ :=\{(x^\alpha_+):z_+>A>0\}$ and
$V^- :=\{(x^\alpha_-):z_-<-A<0\}$, that is $V^+$ is ``on the
right side'' of the original Taub singularity and $V^-$ is ``on
its left side''. We are assuming that the $t,z = const$ surfaces
have the topology of a plane ({\bf R}$^2$) and we can thus assign
to the symmetry $z\rightarrow -z$ a natural meaningful interpretation as
a mirror or specular symmetry. This interpretation seems to be
unwarranted for other choices of topology of the $t,z=const$
surfaces (cf.~Sec.~\ref{sec:discussion}).

The timelike hypersurfaces of junction
$\Sigma^+$ and $\Sigma^-$ are characterized by the
equations $F_+(x^\alpha_+) := z_+-A = 0$ and $F_-(x^\alpha_-):=
z_- +A = 0$, with $A>0$.The induced 3-metrics on $\Sigma^+$ and
$\Sigma^-$ are [cf.\ (\ref{Taub})] identical:
\begin{equation}
ds^2_\pm\mid_{\Sigma^+} = -A^{-2/3}dt_\pm^2+A^{4/3}\left(dx_\pm^2
           +dy_\pm^2\right),    \label{ds2Sigma}
\end{equation}
\noindent thus automatically satisfying the first junction condition.

The verification of the second junction condition requires the
calculation of the extrinsic curvatures of $\Sigma^+$ and $\Sigma^-$.
The unit normal vector to $\Sigma$ (directed from $V^-$ to $V^+$) will have
components $n_+^\alpha$ and $n_-^\alpha$, relative to the charts
$x_+^\alpha$ and $x_-^\alpha$, given by 
\begin{equation}
         n_\pm^\alpha = +\delta^\alpha_3.     \label{n}
\end{equation}

For an intrinsic basis of tangent vectors to $\Sigma$, we will choose
three orthonormal vectors {\bf e}$_{i}$ ($i=0,1,2$) of components
$e^\alpha_{i+}$ and $e^\alpha_{i-}$, relative to the charts
$x_+^\alpha$ and $x_-^\alpha$, given by
\begin{equation}
     e^\alpha_{0\pm}= |A|^{1/3}\delta^\alpha_0,\ \ \ e^\alpha_{1\pm}=
                     |A|^{-2/3}\delta^\alpha_1 ,\ \ \ e^\alpha_{2\pm}=
                                          |A|^{-2/3}\delta^\alpha_2.
\end{equation}

The extrinsic curvature of a hypersurface with unit normal $n^\alpha$
can be calculated as
\begin{equation}
     K_{ij} = e^\alpha_i e^\beta_j n_{\alpha;\beta}, \label{extrinsiccurvature}
\end{equation}
\noindent which furnishes
\begin{equation}
       K^{\pm}_{ij}= \pm\frac{1}{3A}\,\mbox{diag}\,
                            \left(1,2,2\right)
\end{equation}

Lanczos equation \cite{Israel1966}, \(
\gamma_{ij}-g_{ij}\gamma^k_k = - S_{ij} \), where $\gamma_{ij}:=
K^+_{ij}- K^-_{ij}$ is the jump in the extrinsic curvature, then
determines the surface energy-momentum tensor of a shell:
\begin{equation}
       S_{ij} = \frac{2}{3A}\,\mbox{diag}\,\left( -4, 1, 1 \right).
                                        \label{energymomentum}
\end{equation}

\noindent Thus, if we suppose the shell is a perfect fluid one, with proper
four-velocity 
\begin{equation}
      u^\alpha\!\!\mid_\Sigma\, =\!|A|^{1/3}\delta^\alpha_0, 
\end{equation}
\noindent we are forced to recognize that it has a {\em negative\/}
surface energy density. Furthermore, as $A$ approaches $0$, thus realizing the matching ever closer to the
original Taub singularity, we are naturally lead to interpreting
it as an infinite uniform {\em plane\/} distribution of negative
diverging surface energy density. This interpretation is
consistent with the repulsion of geodesics found in
Sec.~\ref{sec:geodesics} and the limit for infinite negative
mass.

\section{Discussion} \label{sec:discussion} 
\setcounter{equation}{0}

In this paper, we carried out an extensive investigation of the
local and global properties of Taub's (static) geometry, and by
extension, of Kasner's (spatially homogeneous) geometry.  First,
we studied the geodesic motion, by using the method of the
effective potential. We showed, in a manifestly
coordinate-independent way, that massive ($\epsilon=1$) free
particles eventually recede from the singular locus, no matter
what initial conditions are attributed to them. Only massless
($\epsilon=0$) particles can attain the singularity, even so
provided their motion is purely ``perpendicular''
($p_{\mid\mid}=0$).  This enlarges upon some results in the
literature\cite{Aichelburg1970}. Second, we showed that Taub's
metric is one of the limits of a family of
Schwarzschild's spacetimes, which is closed in the sense of
Geroch\cite{Geroch1969}. This limit corresponds to the mass
parameter approaching infinitely negative values. This analysis
is carried out by means of Cartan scalars, which characterize in
an invariant way a given geometry. Third, we built
several new plane-symmetric models, by conveniently matching
Taub, Rindler (Minkowski) and Kasner domains.  Fourth, we also
uncovered the asymptotic structure (null, timelike and spacelike
infinities) of Taub's solution.  Fifth, we obtained the complete
weak field limit in asymptotically flat regions, proving that,
at spatial infinity, we may have a nonrelativistic approximately
homogeneous configuration; however, this cannot be associated to
a truly Newtonian limit due to the absence of any essential
parameters which could be related to the mass density of the
source or, in other words, Taub's solution is a single one, not a
family of solutions (like Schwarzschild's one, for instance). We will now take
the opportunity to make some generic comments on related current
literature.

In a recent paper\cite{Bonnor1990}, Bonnor
discusses the difficulties which arise in the interpretation of
the solutions of Einstein's field equations, due to the coordinate
freedom to describe the metric.  Although we agree with this
general idea of his papers, we do not believe that his preferred
(semi-infinite line mass) interpretation of Taub's vacuum static
plane symmetric metric is the {\em only\/} tenable one. Indeed, locally
isometric manifolds can be extended to global manifolds with
distinct topological properties;
this extension cannot be fixed by purely local (in a given
coordinate system) considerations (isometries, geodesics, etc.),
which are doomed to inconclusiveness. For instance, the $t,z =
const$ flat maximally symmetric surfaces of (\ref{Taub}) can be
conceived of as immersed in {\bf R}$^3$ either as topological
planes or topological cylinders, which are, of course, locally
indistinguishable. What does really seems promising, from the
physical point of view, to settle the issue of interpretation is
the realization of experiments which probe the large-scale
structure of the spacetime. Another criterion, one of
simplicity, we have advanced is the {\em mirror-symmetric\/}
matching of Sec.~\ref{sec:junction}, which reproduces the global
Taub solution in the limit of infinitely negative surface energy
density of the plane shell. In short, it seems to us that the
nature of the source, from a purely mathematical point of view,
is a matter of {\em consistent  choice\/} of topology.

We should not conclude without recalling an elegant heuristic
argument of Vilenkin\cite{Vilenkin81}, which might explain the
unreasonableness of looking for a physically viable model for an
infinite plane of constant positive surface density $\sigma$.
Consider, in the alleged plane, a disk of radius $R$, which
contains the total mass  $M(R) = \pi R^2 \sigma$; thus, above a
critical radius, given by $R_c = 1/(2\pi\sigma$, the disk will have
more mass than its corresponding
Schwarzschild mass, and it should collapse (incidentally, the
same argument does not hold for a line or string!). Of course,
compelling as this argument seems, its validity  rests on, at
least, two tacit assumptions: (i)~the intrinsic geometry of the
plane is Euclidean, so that the disk's area is $\pi R^2$,
(ii)~the {\em hoop conjecture\/}\cite{Thorne72}, since we are dealing with a
nonspherically symmetric system. Vilenkin's argument might then
explain why all the proposed
models in GR for an infinite homogeneous plane end up with either a
flat space solution with an appropriate topology or a constant
distribution of {\em negative\/} mass.  In fact, in the literature,there
have appeared two distinct proposals for the general
relativistic problem of a static, infinite, uniformly dense
plane. The first one\cite{Taub1951} is locally isometric to Taub's
plane-symmetric static vacuum spacetime; the second
one\cite{Schucking85,Lemos88} is a Rindler domain, locally isometric
to the Minkowski spacetime. If we insist that a plane massive
configuration may exist in GR, we must cope with the possibility
of negative mass configurations, with an exterior solution given
by Taub's geometry.

As a last remark, one might be tempted to consider the models
inspected in this paper as purely academic ones; however,
already in nonquantum gravitational theory, the issue of
negative mass has a longstanding history, of
which a relativistic landmark is afforded by Bondi's classic
paper\cite{Bondi1957}. Also, from a quantum point of view, it
was early recognized that the weak energy condition cannot hold
everywhere\cite{Epsteinetal1965}. Nowadays, this result has
acquired a renewed interest because, besides the already
mentioned cosmological interest of topological
defects\cite{Peebles1993}-\cite{Coulsonetal1996}, of its import
to the possibility of construction of time machines
\cite{Morrisetal1988} and avoidance of singularities
\cite{Roman1988}. Thus, the best policy seems to be keeping,
{\em cum grano salis\/}, an open mind to these exotic models.

\section*{Acknowledgments}

The authors would like to acknowledge W. Unruh for fruitful
discussions on a preliminary version of this paper, and B.
Mashhoon and G. Matsas for calling attention to some relevant
references. FMP and MOC
acknowledge financial assistance from CNPq.

\appendix

\section{Appendix: the generalized Schwarzschild family}
\label{CS}
\setcounter{equation}{0}

We study a two-parameter family of metrics which includes Taub's and
Kasner's ones and its limits when some parameters are taken to be $0$,
$+\infty$ or $-\infty$, extending the results mentioned in the
last paragraph of
Subsec.~\ref{sec:geodesics}. The formalism is that of the Cartan
scalars, which are the components of the Riemann tensor and its
covariant derivatives calculated in a constant frame and which  provide
a complete local characterization of spacetimes
\cite{Cartan1951}--\cite{Paiva1993}. Spinor components are
used \cite{Penrose1960} and the relevant objects here are the Weyl
spinor $\Psi_A$ and its first and second symmetrized covariant
derivatives $\nabla\Psi_{AB'}$ and $\nabla^2\Psi_{AB'}$. {\sc sheep}
and {\sc classi} \cite{Frick1977,Aman1987} were used in the
calculations.

We shall call ``the generalized Schwarzschild family'' the vacuum 
metrics
\begin{equation} \label{ds2}
ds^{2} = A dt^{2} - A^{-1} dr^{2}
       - r^{2}\left(d\theta ^{2} + K^2(\theta) d\varphi ^{2}\right);  
\ \ A = \lambda - \frac{2m}{r}, 
\end{equation}
where $\lambda$ and $m$ are constant parameters. In the Lorentz tetrads 
($\eta_{AB} = {\rm diag}\,(+1,-1,-1,-1)$)
\begin{equation}  \label{frame}
\begin{array}{lllll}
\theta^{0} = A^{1/2}     dt, &
\theta^{1} = A^{-1/2}    dr, &
\theta^{2} = r           d\theta, &
\theta^{3} = Kr          d\varphi, & 
A > 0  \\
\theta^{0} = (-A)^{-1/2} dr, &
\theta^{1} = (-A)^{1/2}  dt, &
\theta^{2} = r           d\theta, &
\theta^{3} = Kr          d\varphi, & 
A < 0       
\end{array}
\end{equation} 
the nonvanishing components of the Ricci tensor are $R_{22} =
R_{33} = - r^{-2}\left( \lambda + K_{,\theta\theta}/K \right)$.
It can be easily shown that the curvature scalar of the section $t$ and
$r$ constant is given by $2\lambda/r^2$. 
Imposing the {\em vacuum condition\/},
$\lambda = - K^{-1} K_{,\theta\theta}$,
and using the null tetrads
\begin{equation} \label{nullt}
\omega^{0} = \frac{\theta^{0} + \theta^{1}}{\sqrt{2}},~~
\omega^{1} = \frac{\theta^{0} - \theta^{1}}{\sqrt{2}},~~
\omega^{2} = \frac{\theta^{2} + i \theta^{3}}{\sqrt{2}},~~
\omega^{3} = \frac{\theta^{2} - i \theta^{3}}{\sqrt{2}},
\end{equation}
the algebraically independent Cartan scalars are
\begin{eqnarray} 
\Psi_{2} &=& -\frac{m}{r^3} \label{EC}  \\
\nabla\Psi_{20'} &=&  \frac{3}{\sqrt{2}}\frac{m}{|m|}\Psi_2^{\frac{4}{3}}
  \sqrt{\pm\left(\lambda m^{-\frac{2}{3}} + 2\Psi_2^{\frac{1}{3}}\right)} 
\nonumber \\
\nabla\Psi_{31'} &=& \mp \nabla\Psi_{20'}  \nonumber \\
\nabla^2\Psi_{20'}&=&
  \frac{4}{3} \frac{(\nabla\Psi_{20'})^2}{\Psi_2} 
\nonumber \\
\nabla^2\Psi_{31'} &=&  \mp \nabla^2\Psi_{20'}-\frac{3}{2} \Psi_2^2  
\nonumber \\
\nabla^2\Psi_{42'} &=& \nabla^2\Psi_{20'} \nonumber
\end{eqnarray}
(the upper and lower signs corresponding to $A>0$ and $A<0$,
respectively) where, by using the expression  of $\Psi_2$ in
(\ref{EC}), $r$ was eliminated from the other Cartan scalars and, by
using the resulting expression for $\nabla\Psi_{20'}$, we eliminated
$\lambda m^{-\frac{2}{3}}$. From this set of Cartan scalars, one infers
that the metric is Petrov type D, has a one-dimensional isotropy group
given by spatial rotations on the $\omega^2$ -- $\omega^3$ plane (i.e.,
the $\theta$ -- $\varphi$ surface), the isometry group has a
three-dimensional orbit (note that the coordinates $t$, $\theta$ and
$\phi$ do not appear in the expressions of the Cartan scalars) and,
therefore, the isometry group is four-dimensional. 

Since $K(\theta)$ is not present in the Cartan scalars, once $\lambda$
is given and a solution for $K(\theta)$ of the corresponding vacuum condition  is found, any other solution with the same $\lambda$ can
be transformed in this first one by a suitable coordinate
transformation.  Thus, the line element is determined by the values of
$\lambda$ and $m$ only. Moreover, since $\Psi_2$ depends on the $r$
coordinate, by a suitable coordinate transformation, its dependence on
$m$ can be eliminated.  Thus, a member of this family of metrics is
characterized by the sign of $m$, i.e., ${m}/{|m|}$ and the
combination $\lambda m^{-\frac{2}{3}}$ of $\lambda$ and $m$.
Therefore, one can always make a coordinate transformation such that
$\lambda$ becomes $1$, $0$ or $-1$ and $K = \sin\theta$, $1$ or
$\sinh\theta$, respectively. Accordingly, the generalized Schwarzschild
family may be divided into the following sub-families:

\fbox{1) $\lambda > 0$} 
One can make $\lambda = 1$ and $K(\theta)=\sin\theta$, corresponding to
the Schwarzschild line element with $-\infty < m < \infty$.

\fbox{2) $\lambda < 0$} 
One can make $\lambda = -1$ and $K(\theta)=\sinh\theta$. This we shall
call the anti-Schwarzschild line element with $-\infty < m < \infty$.

\fbox{3) $\lambda = 0$} 
One can make $K = 1$. The factor $\lambda m^{-\frac{2}{3}}$ becomes
zero. Therefore the only parameter is the sign of $m$ and its absolute
value can always be made equal to $1/2$ (for $m = 0$ the line element
becomes singular).

\fbox{3a) $\lambda = 0$ and $m > 0$} One can make $m = 1/2$ leading to
a single metric, namely Kasner's one, (\ref{Kasner-r}). From
(\ref{EC}), the Cartan scalars are
\begin{equation}\label{ECKasner}
\begin{array}{l}
\Psi_{2} =  -\frac{1}{2r^3},\ \  
\nabla\Psi_{20'} = 3\Psi_2^{\frac{3}{2}},\ \
\nabla\Psi_{31'} = \nabla\Psi_{20'},\\ 
\nabla^2\Psi_{20'} = - 12\Psi_2^2,\ \ 
\nabla^2\Psi_{31'} = -\frac{27}{2}\Psi_2^2,\ \
\nabla^2\Psi_{4 2'} = \nabla^2\Psi_{20'}\ .
\end{array}
\end{equation}

\fbox{3b) $\lambda = 0$ and $m < 0$} One can make $m = -1/2$ leading to
a single metric, namely Taub's one, (\ref{Taub-r}). From
(\ref{EC}), the Cartan scalars are
\begin{equation} \label{ECTaub}
\begin{array}{l}
\Psi_{2} =  \frac{1}{2r^3},\ \  
\nabla\Psi_{20'} = -3\Psi_2^{\frac{3}{2}},\ \
\nabla\Psi_{31'} = - \nabla\Psi_{20'},\\  
\nabla^2\Psi_{20'} = 12\Psi_2^2,\ \
\nabla^2\Psi_{31'} = -\frac{27}{2}\Psi_2^2,\ \
\nabla^2\Psi_{4 2'} = \nabla^2\Psi_{20'}\ . 
\end{array}
\end{equation}

The last two line elements are special cases of the Kasner-type
metric \cite{Harvey1990}, $ds^2 = et^{2a_1}dt^2 - et^{2a_2}dx^2 -
t^{2a_3}dy^2 - t^{2a_4}dz^2$, where $a_2 +a_3 + a_4 = a_1 + 1$,
$(a_2)^2 + (a_3)^2 + (a_4)^2 = (a_1 + 1)^2$ and $e = \pm 1$.  To
recover (\ref{ECKasner}) we may choose $a_1=\frac{1}{2}$,
$a_2=-\frac{1}{2}$, $a_3=a_4=1$ and $e = +1$; and to recover
(\ref{ECTaub}) we may choose $a_1=\frac{1}{2}$, $a_2=-\frac{1}{2}$,
$a_3=a_4=1$ and $e = -1$.

Therefore, the generalized Schwarzschild family (\ref{ds2}) may be
divided into two one-parameter families, namely the Schwarzschild
family and the anti-Schwarzschild family plus the Kasner metric
(\ref{Kasner-r}) and the Taub metric (\ref{Taub-r}). The analysis of
the Cartan scalars shows some interesting features of the parameters
and functions appearing in the original line element (\ref{ds2}). As
concerns Taub's metric, we showed that it is a particular case of the
generalized Schwarzschild family with $\lambda = 0$ and $m < 0$.
Although the parameter $m$ is still arbitrary, it can be absorbed away
by a coordinate transformation and Taub's geometry is thus independent
of parameters: it is a {\em single\/} solution, not a family. A
consequence of this fact emerges in the weak field limit  (\ref{Phi}),
which has no essential parameter.

From the Cartan scalars (\ref{EC}), we see that, as
$m\rightarrow\pm\infty$, the limits depend only on the sign of $m$ and
not on $\lambda$, therefore the Schwarzschild and the
anti-Schwarzschild families have the same limits. The limiting
procedure we adopt is that of \cite{Paivaetal1993}, where
we choose limits for $\Psi_2$ and find the limits of the other Cartan
scalars. The possible limits for $\Psi_2$ are (i) $0$, (ii) a nonzero
constant and (iii) an arbitrary function of the coordinates. (i) If
$\Psi_2 \rightarrow 0$, then all Cartan scalars (\ref{EC}) tend to
zero, therefore both families tend to Minkowski spacetime (which is
Petrov type 0). (ii) As discussed by \cite{Paivaetal1993},
Petrov type D metrics cannot have $\Psi_2$ constant while the other
components of the Weyl spinor are zero. (iii) If $\Psi_2$ tends to an
arbitrary function of the coordinates, then as $m \rightarrow +\infty$
both families tend to the Kasner metric (\ref{Kasner-r}) and as $m
\rightarrow -\infty$ both families tend to the Taub (\ref{Taub-r}) as
can be easily seen comparing the the Cartan scalars obtained in the
limit with those of Kasner (\ref{ECKasner}) and Taub (\ref{ECTaub}). As
discussed in \cite{Paivaetal1993}, different functional
forms of the limit of $\Psi_2$ do not lead to different limits, since
coordinate transformations can take one form to the others. This
analysis covers all Petrov type D and 0 limits.

Geroch \cite{Geroch1969} introduced the concept of a closed family of
metrics, i.e., a family that contains all its limits. Considering the
family of metrics defined by (\ref{ds2}), we can say that it is closed
under Petrov type D and 0 limits as $m$ tends to $+\infty$ or $-\infty$.

As a matter of completeness, we mention that the limits of the
generalized Schwarz\-schild family as $\lambda \rightarrow 0$ and $m > 0$
or $m < 0$ are exactly the same as the limits as $m\rightarrow \pm
\infty$. This can be easily seen from the Cartan scalars
(\ref{EC}), noting specially the expression of $\nabla\Psi_{20'}$.

\newpage
\section*{Table and its caption}

\begin{table}[h]
\begin{tabular}{|l|l|l|l|}
\hline 
coord.      & definition               & range                 & 
                                          geometrical loci \\
\hline
$tzxy$      &                          & $0\leq z < +\infty$   & 
                                       $\ast$       : $z=0$ \\
\hline
$trxy$      & $r=(3z/2)^\frac{2}{3}$   & $0\leq r < +\infty$   & 
                                       $\ast$       : $r=0$ \\ 
\hline
$uvxy$      & $u=e^{-r^2/4}\sinh(t/2)$   & $0\leq v < +\infty$   & 
                                 $\ast$       : $v^2-u^2=1$ \\
            & $v=e^{-r^2/4}\cosh(t/2)$   & $0\leq v^2-u^2 \leq 1$   & 
                                    ${J^+}$ : $v=+ u\not=0$ \\
            &                          &                       & 
                                    ${J^-}$ : $v=- u\not=0$ \\
            &                          &                       &
$i_0$        : $u=0,\ v=0_+$ \\
\hline
$\mu\nu xy$ & $\mu = -\ln(v-u)$        & $0\leq(\mu+\nu)<+\infty$ & 
                             $\ast$       : $\mu + \nu =0 $ \\
            & $\nu = -\ln(v+u)$        &                       &
${J^+}$ : $\mu\rightarrow+\infty$,\ $\nu$ \mbox{ finite} \\
            &                          &                       &
${J^-}$ : $\nu\rightarrow+\infty$,\ $\mu$ \mbox{ finite} \\
            &                          &                       & 
     $i_0$        : ($\mu\rightarrow+\infty$, $\nu\rightarrow+\infty$) \\
\hline
$\xi\nu xy$ &$\xi=\sqrt{4\mu}$, $\mu>0$& $0 < \xi < +\infty$ \\
\cline{1-3}
$\xi NXY$   & $X=\xi x$, $Y=\xi y$                             \\
            & $N = \nu - \xi(x^2+y^2)$                         \\
\cline{1-2}
$TZXY$      & $Z = (N+\xi)/2$                                  \\
            & $T = (N-\xi)/2$                                  \\
\cline{1-2}
\end{tabular}
\caption{Coordinate systems used for Taub metric in this paper.
Homonymous coordinates may differ by a constant and unspecified
ranges are $(-\infty,+\infty)$. An $\ast$ in the fourth column means
the singularity. The last two coordinate systems are meaningful only in
the asymptotic region. For the Kasner metric, we interchange $u$ and
$v$ in the table; the point $(u=0_+,v=0)$ for the Kasner domain now
represents the future timelike infinity $i^+$, instead of $i^0$.
Replicas of these spacetimes (K and T) in the $(u,v)$ plane can be
obtained by changing $(u,v)\rightarrow (-u,-v)$.}
\label{coordinates}
\end{table}

\newpage
\section*{\bf Figure captions}

\begin{figure}[h]
\caption[]{The Taub spacetime, (\protect\ref{ds2t}), with two coordinates $(x,y)$
suppressed. Each point of the diagram corresponds to a 2-dim spacelike
plane. The spacetime manifold is the quasi-compact region of the
$(u,v)$ plane bounded by the timelike singularity $v^2-u^2=1$, the null
boundaries $J^+$ and $J^-$, and the spatial infinity $i^0\
(u=0,v=0_+)$. The null geodesics (with $p_{||}=0$) are forty-five
degrees straight lines. With respect to the metric (\ref{ds2t}) $u$ and
$v$ are timelike and spacelike coordinates, respectively.} \label{I1}
\end{figure}

\begin{figure}[h]
\caption[]{The Kasner spacetime, (\protect\ref{ds2k}), with coordinates $(x,y)$
suppressed. The future flat null infinities $J^+$ correspond to
$u^2-v^2=0$. The $(u,v)$ coordinates are now related to the $(t,r)$
coordinates of (\protect\ref{Tds2}) with $-\infty < r \leq 0$ by $u =
f(r)\cosh(t/2)$, $v = f(r)\sinh(t/2)$, in order that
$(u,v)$ have timelike and spacelike character, respectively. The
future timelike infinity $i^+$ is the point $(u=0_+,v=0)$.}
\label{I2}
\end{figure}

\begin{figure}[h]
\caption[]{The relation between the $(u,v)$ and $(\mu, \nu)$ coordinates.
The gray region of (b) corresponds to the gray T region in (a) bounded by
the flat null infinities $J^+$ and $J^-$, the null surfaces
$\nu = 0$ and $\mu = 0$ plus the singular point  $\mu = 0 = \nu$.
Coordinate systems may be introduced such that on the null surfaces
$\mu = 0 $ and $\nu = 0 $ the metrics $ds^2_K$ and $ds^2_T$ have the
form of Minkowski metric in Cartesian coordinates, except at the
singular point $\mu=0=\nu$.}
\label{I4}
\end{figure}

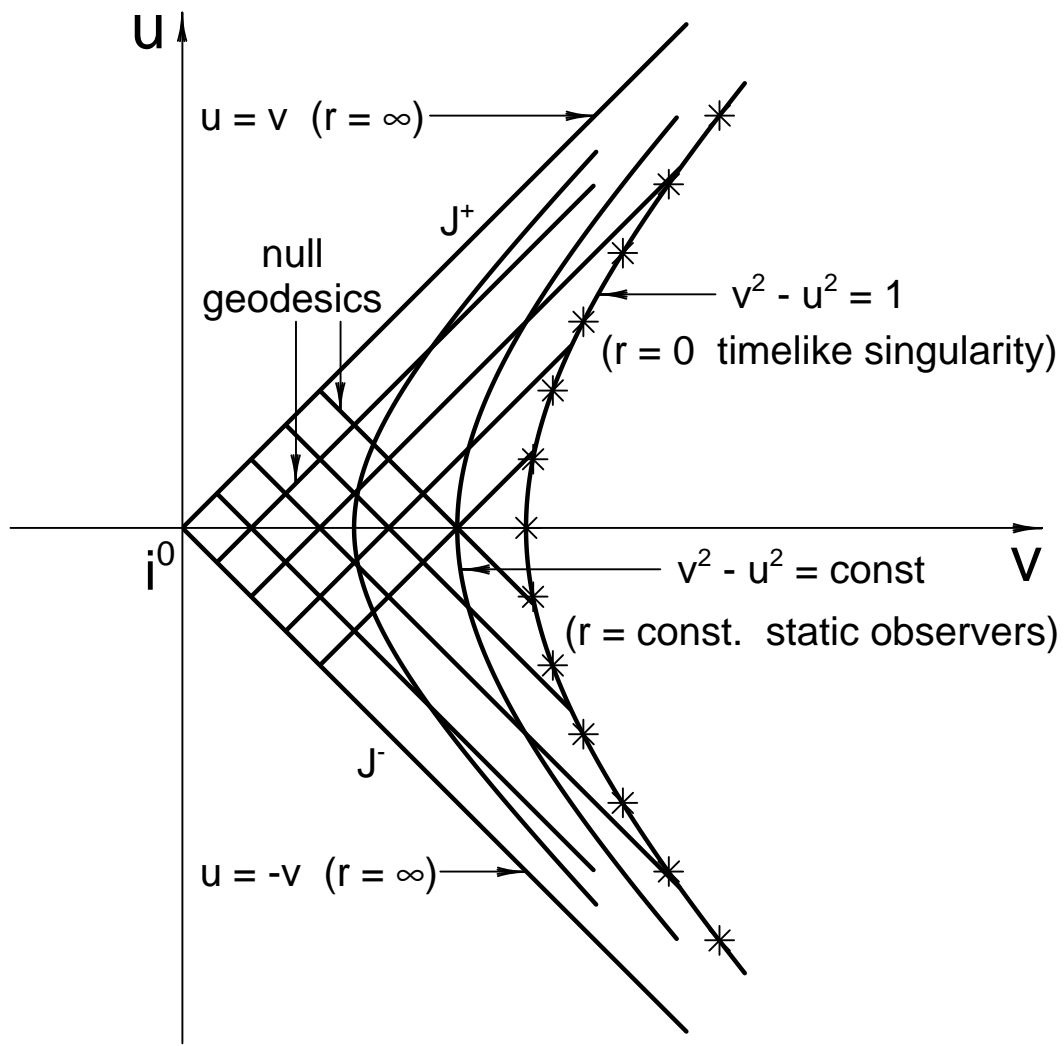
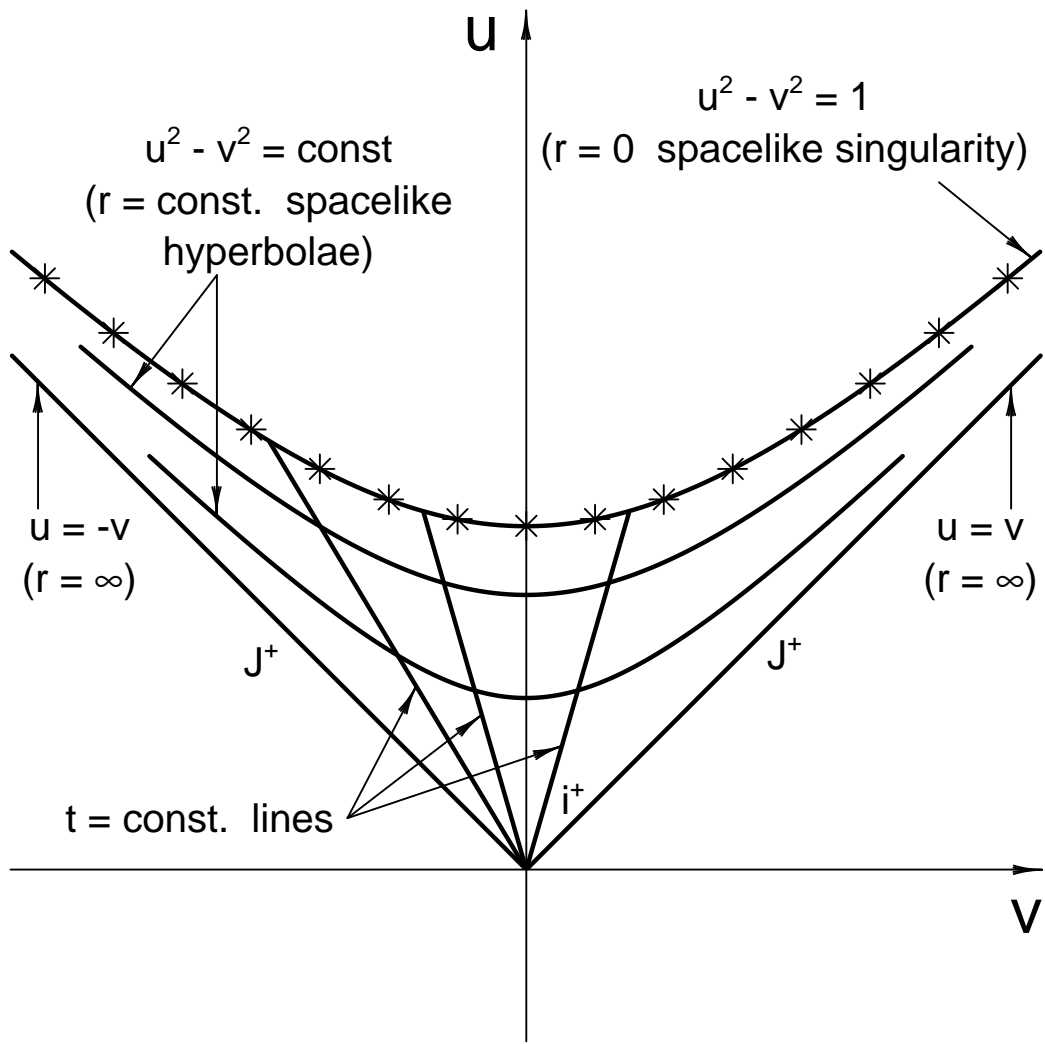
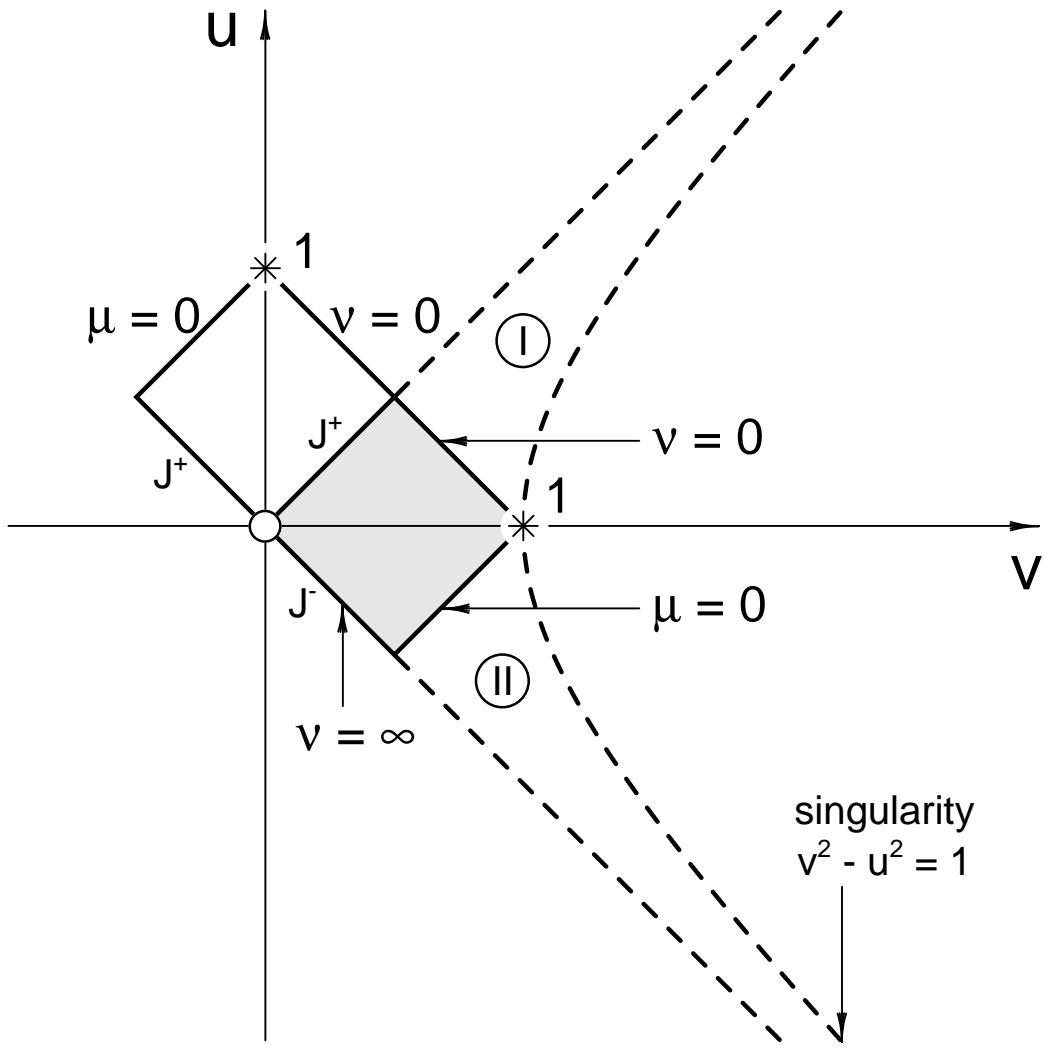
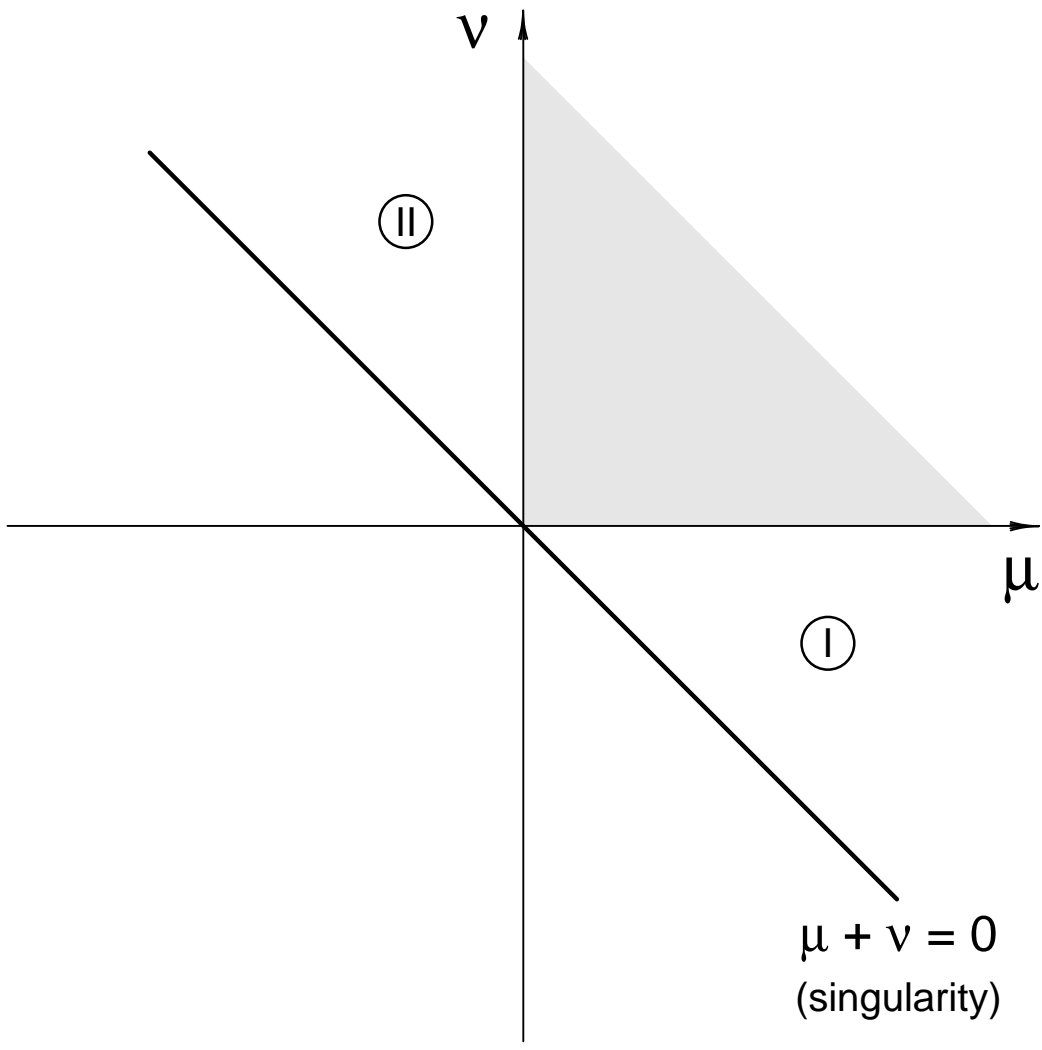
\end{document}